# Gamma-Ray Line Emission from Superbubbles in the Interstellar Medium: The Cygnus Region

S. Plüschke[1], R. Diehl[1], K. Kretschmer[1], D. H. Hartmann[2], U. Oberlack[3]

[1] *MPE Garching, Giessenbachstrasse, 85748 Garching, Germany*
[2] *Dep. of Physics & Astronomy, Clemson Univ., Clemson, SC 29634, USA*
[3] *Astrophys. Lab. Columbia Univ., New York, NY 10027, USA*

**Abstract.** The star forming process in the Milky Way is non-uniform in time and space. The scale of star forming regions ranges from groups within a few pc to large segments of spiral arms with linear dimension of order kpc. When many stars form in a relatively small volume over a short duration, a localized starburst ensues. The energetic impact of such a burst of star formation can severely affect the dynamic structure of the gaseous disk. Stellar winds and supernova explosions drive an expanding superbubble, whose size eventually exceeds the scale height of the disk and thus drives a disk-wind blowing metal enriched gas into the halo. We discuss the basic scenario of superbubble evolution, emphasizing the associated gamma-ray line signatures. In particular, we discuss nuclear line emission from $^{26}$Al and $^{60}$Fe in the Cygnus region.

## INTRODUCTION

The COMPTEL experiment aboard the Compton Observatory has produced a nine year all sky survey in the MeV regime [18]. Here we emphasize the all sky map in a narrow band centered on 1.809 MeV which selects the radioactive decay line from $^{26}$Al (Figure 1). With a mean lifetime of roughly 1 Myr there are many individually unresolved sources which contribute to this diffuse glow of the Galaxy (for a recent review see [17]). The bulk of the observed flux is due to $^{26}$Al emerging in the winds of massive stars and the ejecta of core-collapse supernovae with relative contributions of 60% and 40%, respectively [8]. Nucleosynthesis of $^{26}$Al in Novae and AGB stars also contributes to the glow, but probably to a lesser extent. Population synthesis studies suggest that the galactic disk presently contains about 2 $M_\odot$ of radioactive $^{26}$Al. In steady-state a global galactic star formation rate of a few solar masses per year corresponds to a supernova rate of a few events per century, which implies a typical yield of $10^{-4}$ $M_\odot$ of $^{26}$Al per supernova. This observationally determined yield is consistent with the theoretical values derived from hydrostatic and explosive nucleosynthesis simulations (e.g. [11,23]).

Multi-wavelength observations of the Milky Way and external galaxies clearly show that star formation exhibits a hierarchical pattern. Most stars form in groups, associations, clusters, and still larger conglomerates [4]. Thus, star formation is is a strongly correlated process. The transformation of $10^4$ to $10^5$ solar masses of gas into stars (with a conversion efficiency of roughly 10%) leads to a subsequent dynamic epoch due to the winds and explosions from several hundred stars in the 8 to 100 $M_\odot$ range. The kinetic energy input into the surrounding interstellar medium from these stars drives the expansion of a "super-remnant" which can even blow a "hole" into the gaseous disk. There is much observational evidence for such superbubbles in the Milky Way and other nearby galaxies. Superbubbles provide the means by which the disk and the halo are chemo-dynamically coupled. The interior of a superbubble, similar to a supernova remnant, contains a tenuous, high temperature gas. If the $^{26}$Al ejected by massive stars attains high velocities in this environment we may expect significant broadening of the observed 1.8 MeV line. In fact, the GRIS experiment has measured a FWHM of 5.4±1.2 keV, which corresponds to temperatures somewhat higher than $10^8$ K [12]. Temperatures in this regime are actually very hard to accomplish in the ISM and also hard to maintain on timescales comparable to the mean life of $^{26}$Al. Alternatively, the radioactive aluminum nuclei could be embedded in high velocity dust grains (perhaps accelerated in superbubble interiors), so that the observed line broadening is not thermal in nature but a purely kinematic effect. In that case typical velocities should be of order 500 km/s. Assuming free expansion these velocities would lead to an angular scale height that exceeds the COMPTEL observations. It is currently unclear if a broad 1.8 MeV line is a global feature, or if only localized regions on the sky exhibit this effect while others show less or perhaps no broadening. This issue is likely to be resolved with data from the INTEGRAL mission [22].

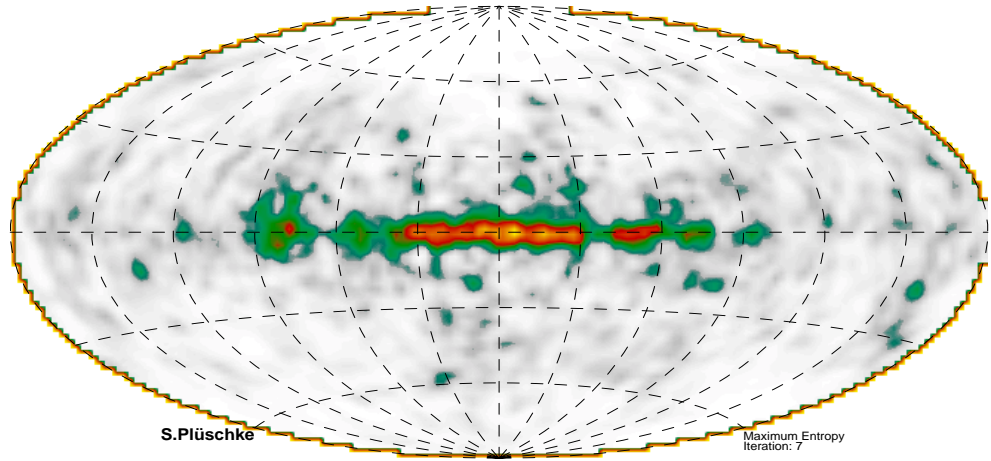

**Figure 1.** Galactic map at 1.8 MeV obtained with COMPTEL [15]. The prominent extended emission feature between l = 70° and l = 90° coincides with the Cygnus star forming region.

Here we focus on one particular star forming region in the Galaxy, the Cygnus region. We model the observed 1.8 MeV flux with a variety of OB associations whose properties are discussed in the literature. We also provide a flux estimate at 1.173 and 1.332 MeV from $^{60}$Fe.

# MODELLING THE CYGNUS REGION

The Cygnus region (defined by the extended 1.8 MeV emission feature at $l = 80\pm 10$ deg, see Fig. 1) contains numerous massive stars. The galactic O star catalogue [5] lists 96 O stars in this field. In addition, there are 23 Wolf-Rayet stars of which 14 are of WN-type, 8 of WC-type, and 1 is classified as WO [21]. Between 5 and 10 of these stars are believed to be members of OB associations. In addition, the Galactic SNR Catalogue lists 19 remnants in this region. For 9 of these remnants reliable age and distance estimates are available [7]. In our model WR stars and SNRs that do not belong to one of the Cygnus OB associations are treated as isolated sources whose estimated 1.8 MeV flux is subtracted from the integrated COMPTEL measurement. The resulting residual flux is shown in Fig. 3.

In the Cygnus region one finds numerous open clusters and nine OB associations [1,6,10]. Two of those (OB 5 & 6) could be unrelated due to projection and selection effects [6]. Cygnus OB4 has poorly determined properties. Therefore, the remaining six OB associations (1,2,3,7,8,9) were chosen as a basis set for our model.

Optical studies of the Cygnus region are hampered by a giant molecular complex along the line of sight [3]. From the 2MASS NIR survey of a field centered on Cygnus OB2 the estimated number of O stars has tripled [9]. Extinction corrections in this region are severe. Knödlseder [9] inferred $120\pm20$ O stars and $2500\pm500$ B stars. Similar corrections are expected for the remaining OB associations in Cygnus. We used the CO map [3] to estimate the extinction correction factors. The procedure was calibrated with observations in the optical and NIR of Cyg OB2.

Based on the observed numbers of O stars, corrected for extinction, we generate synthetic stellar populations for each association by means of Monte Carlo sampling of an appropriate initial mass function. MC sampling allows us to derive statistical uncertainties due to the relative small size of the associations. Our population synthesis model (e.g. [13,14]) provides the time evolution of the interstellar masses of $^{26}$Al and $^{60}$Fe, the kinetic power due to stellar winds and supernova explosions, and the emission of ionizing photons. We also determine the number of WR stars as function of time. Adding these WR stars from each association the expected total number of WR stars potentially observable is of order $10\pm3$. This estimate is consistent with the reported number of associated WR stars in the Galactic WR Star Catalogue [21].

Gamma-ray lines from the decay of $^{26}$Al and $^{60}$Fe follow directly from the predicted mass budget. Uncertainties in flux predictions are due to a combination of statistical errors (small population size), systematic uncertainties (e.g. stellar yields), and the often poorly known distances to the associations [14].

The release of freshly synthesized nuclei is directly coupled to the injection of kinetic energy into the interstellar medium. Typical stellar wind power is $\sim 10^{37}$ erg/s and supernovae deliver $\sim 10^{51}$ erg of kinetic energy to the surrounding medium (but see [20]). Thus OB associations drive large cavities into the interstellar medium. The interior of these cavities (superbubbles) is filled with a hot, tenuous plasma, and their size can reach several 100 pc. Because of the high ejection velocities (several 1000 km/s) and the low density in the cavity the freshly ejected nuclei travel large distances before they decay. Assuming a mean velocity of 1000 km/s $^{26}$Al nuclei could traverse a distance of ~200 pc within one fifth of their mean lifetime. Therefore the diffuse nature of the observed 1.809 MeV emission has two contributions, the spatial distribution of the many contributing sources and the dynamic expansion.

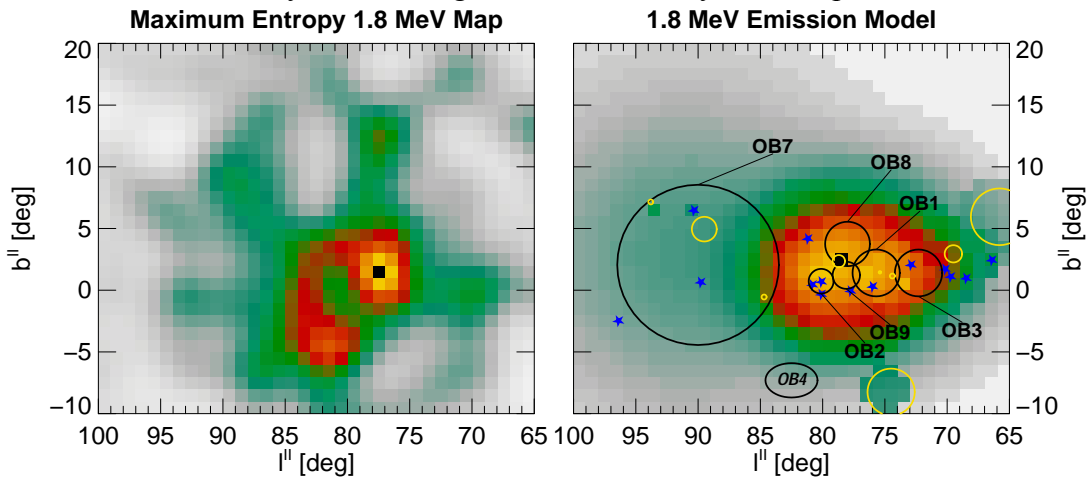

**FIGURE 2.** (left side) Detailed COMPTEL 1.8 MeV map of the Cygnus region. (right side) angular distribution of known OB associations, WR stars and SNRs overlaid on the $^{26}$Al distribution model.

We developed a 1D analytical model of expanding superbubbles [16], similar to the model of [19], to simulate the dynamic evolution of the $^{26}$Al distribution in the ISM. The model assumes a uniform ambient density and drives an expanding thin shell by a central, variable energy source. The time dependent luminosity of the central source is obtained from the population synthesis model. Energy losses due to radiation from the bubble interior are taken into account. We assume that $^{26}$Al fills the bubble interior such that the surface brightness in the $\gamma$-ray line is constant. We compare the model (Fig. 2 right) with the data (Fig. 2 left) by means of Maximum Likelihood fitting. The model fits suggest a mean ambient density between 20 and 30 cm$^{-3}$, consistent with values inferred by other methods (e.g. [2]).

In addition, the fits imply the presence of a low-intensity component across the whole Cygnus region, which could be the result of a small amount of $^{26}$Al in the nearby foreground. It is also possible that several unidentified novae and/or AGB stars explain this low-intensity component. Yet another alternative is a small contribution from the very extended associations Cygnus OB5 & 6 [1], which were not included in this study for reasons given above.

# CONCLUSIONS AND PREDICTIONS FOR $^{60}$FE

Fig. 3 shows that the multiple OB association model for Cygnus successfully describes the observed 1.8 MeV flux. Roughly 70% of the observed flux is due to the OB associations, while about 20% of the flux is attributed to isolated WR stars and SNRs in this region. The origin of the remaining 10% is not yet clear.

In addition, from this model of the massive star population in the Cygnus region we can predict the $\gamma$-ray line fluxes at 1.173 and 1.332 MeV due to radioactive decay of

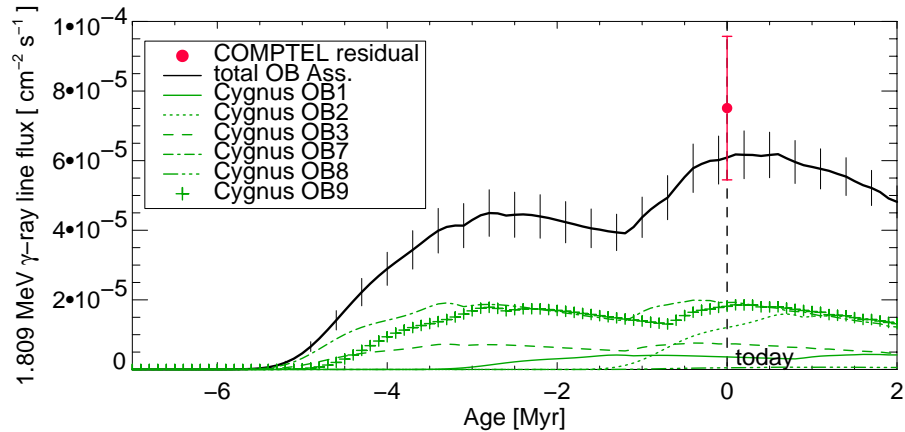

**FIGURE 3.** Flux at 1.8 MeV from $^{26}$Al as a function of time predicted for the Cygnus star forming region. The present day value (Age = 0) is reproduced by the superposition of $^{26}$Al produced by six well known OB associations in this region. The data point shown here represents the integrated flux minus an estimated contribution from isolated, non-associated WR stars and SNRs.

$^{60}$Fe, which is also abundantly produced by massive stars. Our best fit model predicts a total flux of $(9 \pm 5)\ 10^{-6}$ photons cm$^{-2}$ s$^{-1}$ for each line. Even if these lines are combined, a detection with COMPTEL is rather unlikely. The INTEGRAL mission [22] is significantly more sensitive to narrow lines in this energy regime, but a detection of $^{60}$Fe specifically in Cygnus would require a substantial fraction of the available observation time. However, a detection of $^{60}$Fe from the inner Galaxy is indeed expected within the first year of the INTEGRAL mission.